# Photoelectron superlubricity

*Cheng Chen[1], Zhixin Zhang[1], Fan Lei[1], Haifeng Weng[1], Peidong Xue[1],*

*Dongfeng Diao[1,2,3] \**

1. Institute of Nanosurface Science and Engineering, Shenzhen University, Shenzhen 518060, China;

2. National Key Laboratory of Radio Frequency Heterogeneous Integration, Shenzhen University, Shenzhen 518060, China;

3. Electron Microscopy Center of Shenzhen University, Shenzhen 518060, China;

\*Correspondence. E-mail: dfdiao@szu.edu.cn.

**Summary**

Superlubricity, a state where friction between two contact surfaces is nearly zero, has a great potential to revolutionize various mechanical systems by significantly reducing energy dissipation and enhancing efficiency[1-11]. It can be realized either by structural incommensurate contact between crystalline surfaces or by creating highly passive interfaces to cancel out the adhesive forces. However, fabricating and maintaining such superlubric surfaces still present challenges, often disabled by surface structural defects or susceptibility to humid atmospheres[12-16], which renders superlubricity fragile. Here, we propose a novel strategy of photoelectron superlubricity (PESL), where robust superlubricity can be achieved in humid atmospheres by in-situ laser-irradiating the contact interface of an amorphous carbon film sliding against sapphire ball. We demonstrate that PESL not only exhibits a high resistance to environmental disturbances but also features rapid response. The formation of PESL originates from the laser-irradiating induced formation of nanographene-layered interface and enrichment of photoelectrons at the interface, resulting in a repulsive electric field between the nanographene layers. The discovery of PESL opens a new avenue for achieving superlubricity, and also provides novel insights for smart friction, mechanical motion control and light manipulation.

**Main text**

Superlubricity, enabling near-frictionless sliding between surfaces, has significantly advanced the frontiers of friction reduction and control. From the perspective of practical application, it is the most ideal friction situation for many moving mechanical systems mainly because friction consumes large amounts of energy and causes system failure. Nowadays, various superlubricity systems[1-12], especially based on carbon materials, have been discovered and verified. For instance, structural superlubricity can be achieved with perfect graphene layers. Since the prediction of the structural superlubricity theory in the early 1990s[1-2], through to its nanoscopic validation with highly oriented pyrolytic graphite (HOPG) in 2004[3], and subsequent advancements across microscale in 2012[4,5] to macroscale in 2015[6], each breakthrough in this field has been exciting and fascinating.

Generally, achieving structural superlubricity demands that contact crystalline surfaces maintain dry, rigid, and clean to avoid defects, and the atomic or molecular structures of the two surfaces are incommensurate. Structural defects on the contact surfaces, whether inherent or induced during the friction process[12-14], can lead to the failure of superlubricity and are difficult to repair. Therefore, applying the principles of structural superlubricity into macroscopic friction is very challenging. Although high-quality graphene layers can be obtained from HOPG or fabricated through advanced wafer-scale CVD technologies, challenges persist in scaling up to macroscopic contacts while avoiding grain boundary[15] or step[16] structures and ensuring an incommensurate contact state.

Recently, various strategies[17-19] for achieving macroscale structural superlubricity have been proposed by segmenting macroscale sliding interface into multiple microscale incommensurate contact sites. For instance, macroscale superlubricity can be achieved by using graphene-wrapped nanodiamonds sliding against diamond-like carbon surface[6] or by constructing 2D layered heterostructures on textured substrates[17]. However, these superlubricity systems still exhibit certain systemic fragility, particularly in terms of tolerance to structural defects and atmospheric conditions, which are insufficient. Moreover, they lack the capabilities for self-repair and real-time regulation.

Hydrogenated amorphous carbon (a-C:H) films are another typical carbon material that have been extensively studied and widely used for their macroscale superlubricity properties in inert environment[20-22]. The superlubricity of a-C:H films is primarily attributed to the passivation of carbon dangling bonds by hydrogen at the sliding interface. The incorporation of hydrogen into the amorphous carbon (a-C) films, either during the film deposition, through post-processing, or during the friction process itself, can effectively achieve superlubricity, demonstrating good controllability[20].

Regrettably, the superlubricity of a-C:H films also fails in atmospheric conditions due to the susceptibility of a-C:H surfaces to tribochemical reactions with water and oxygen molecules[22].

In brief, realizing superlubricity, whether through structural incommensurate contacts or highly passive interface, is still challenging and often undermined by surface defects and humidity. To address this challenge, we propose an innovative strategy for macroscale superlubricity through in-situ fabricating nanographene-layered interface with electron repulsion. Specifically, we integrated a laser beam into a tribo-system (Fig.1a and Supplementary Fig. S1). The laser irradiation, after passing through a transparent sapphire ball, ensured full coverage of the friction contact area (as shown in the inset of Fig.1a, where the laser spot radius exceeds the contact radius). From Fig. 1b, we can see a dramatic reduction in the friction coefficient to 0.007, reaching the superlubric state. The reduction of the friction coefficient depended on the laser power density (Inset of Fig. 1b). Optical imaging and profilometry revealed that the wear track on the a-C film was nearly undetectable (Fig. 1c), indicating an ultra-low wear characteristic with a remarkably smooth worn surface (Supplementary Fig. S5). While the a-C film showed much more wear when there is no laser irradiation, this result highlights the distinctive property of PESL. Raman spectra revealed that the nanostructure of wear track mainly maintained an amorphous structure, with no significant graphitization observed. The surrounding contact area of the counter ball was covered with a transferred film that had undergone graphitization, as indicated by the increased intensities of the D and G peaks in Raman spectrum. Therefore, the laser irradiation primarily affected the top-surfaces of the friction interface, resulting in ultra-low friction and wear.

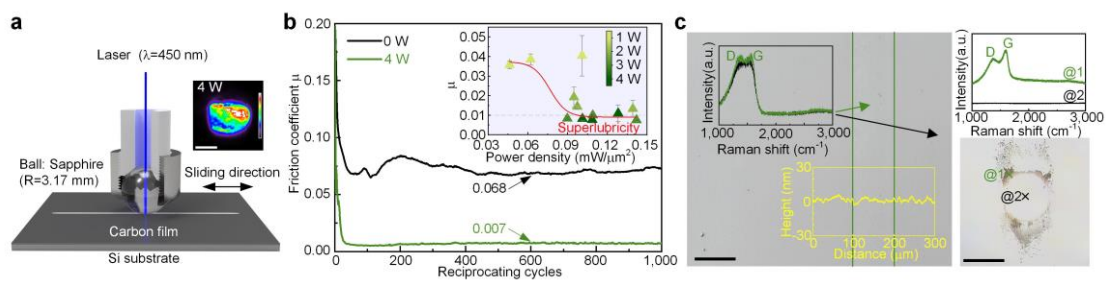

**Fig. 1 | Photoelectron superlubricity. a**, Schematic illustration of PESL test. Inset is the image of 4 W laser spot. **b**, Typical friction coefficient curves of amorphous carbon film sliding against sapphire ball with 0 W and 4 W laser irradiation in atmospheric environment (Temperature: 24 °C; Humidity: 50~70%). Inset is the evolution of friction coefficient with power density. At the same laser power, different power densities can be obtained by adjusting lens position to change the spot size. **c**, After 1000 reciprocating cycles of friction in 4 W superlubricity condition, the wear track of a-C film was almost indistinguishable in optical image and scan profiler, and no significant graphitization occurred as indicated in Raman spectra. A graphitized transfer film was formed, surrounding the contact area of the sapphire ball. The tests were performed under 5 N load and with 20 mm/s linear speed. Scale bars, 1 mm (**a**, inset), 100 μm (**c**). a.u., arbitrary units.

To confirm the generality and uniqueness of PESL, we demonstrate that PESL is achievable across a broad load range of 3 to 13 N (Contact pressure ~1 Gpa) and is maintained in various gas environments such as humid air, nitrogen, and argon (Supplementary Fig. S6). Notably, PESL kept stable under multiple alternating gas, and the friction coefficient could further reduce in inert gas (Fig.2a). This also indicates that atmospheric interference from water and oxygen molecules is consistently present, but laser irradiation can effectively enhance resistance to such environmental disturbances. This robustness of PESL is also evident in the friction response behavior under alternating on/off laser irradiation (Fig.2b). The transition to the superlubric state upon laser on occurred within 17.4 seconds (<9 reciprocating cycles, Fig.2c), and the loss of superlubricity upon laser off was nearly instantaneous (response time <1 s, Fig.2d). The rapid response and good robustness represent the excellent controllability of PESL.

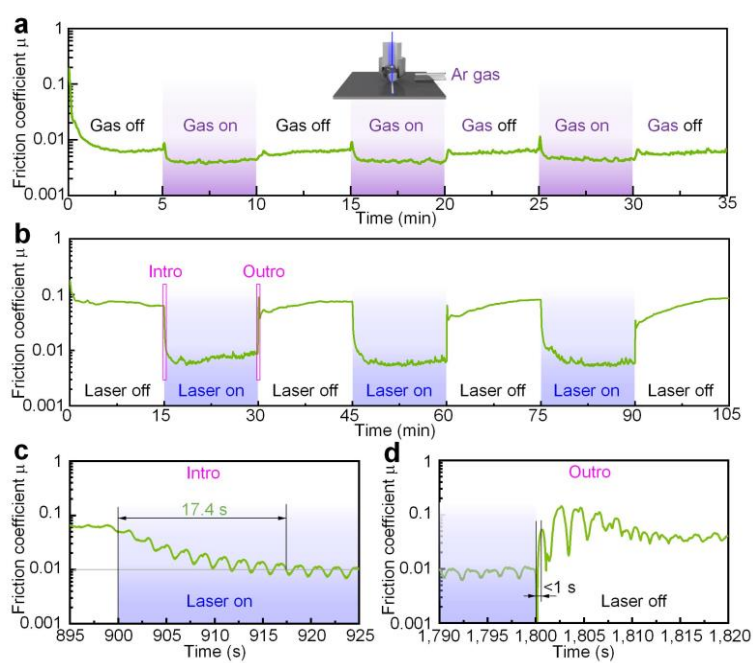

**Fig. 2 | Robust and rapid response of PESL. a**, Friction coefficient curve of PESL in alternating humid air/Ar environment by controlled blowing argon gas. PESL kept stable under multiple alternating gas, and friction coefficient could further reduce in inert gas. **b**, Response friction behavior of PESL under alternating on/off laser irradiation. **c** and **d**, The on/off response times were 17.4 s and <1 s respectively.

To uncover the mechanism of PESL, we observed the nanostructure of the transfer film by transmission electron microscopy (TEM), and found it to be composed of parallel-stacked nanographene layers (Fig.3a). The nanographene is approximately 2~5 nm in size. Electron energy-loss spectra (EELS) confirmed the transfer film is mainly composed of $sp^2$ bonded carbon with a high $\pi^*$ peak (at ~285 eV) in the carbon K-edge. High-magnification image revealed the presence of structural defects both within the inner layers and at the edges of the nanographene (Fig.3b). Correspondingly, we also observed the presence of nanographene layers at the surface of the wear track (Supplementary Fig. S7), even though the wear track is predominantly composed of amorphous carbon, which makes it challenging to directly observe the top-surface nanostructure using TEM from plan view. This in-situ formed nanographene at the sliding interface is benefit to create a natural incommensurate contact. Nevertheless, as is known, structural defects are a primary cause in superlubricity failure, especially in atmospheric environments. Therefore, we deduce that there must be another key mechanism in PESL that shields the adverse effects of structural defects.

Electron-electron interaction plays an essential role in friction at the sliding interface[23-27], which leads us to consider that photo-induced electrons can be instrumental in PESL. To confirm this, the sapphire ball was coated with a thin Au layer to construct a circuit for measuring the current from the friction interface (Fig.3c). We can see that PESL is accompanied by an order-of-magnitude increase in the measured current (Fig.3d), with electrons flowing from the carbon film surface towards the counterball. It is important to note that the measured current likely represents only a fraction of the electrons escaping from the friction contact center (the center of laser irradiation), forming a spontaneous current. Therefore, it is reasonable to believe that the electron density at the contact center was significantly higher.

A vast number of electrons are confined at the friction interface, where they can be trapped by nanographene layers with unsaturated defect or edge structures[28-30], forming a considerable repulsive electric field (Fig. 3e). Photoelectron generation and conduction are very rapid, occuring on femtosecond to nanosecond time scales. So, when the laser was turned off, the loss of photo-induced electrons and the concomitant repulsive electric field was instantaneous, leading to an immediate increase in friction force, even though the nanographene layers remained on the surface.

Additionally, from this perspective, we can also observe that the system did not transition to a state of superlubricity immediately upon laser activation. This delay suggests that the repulsive electric field established by photoelectrons alone is insufficient to induce superlubricity. Instead, a certain period is required for facilitating the formation of nanographene layers by friction and laser irradiation. PESL is thus believed to arise from the synergistic effects of laser-induced nanographene layer formation at the interface and the enrichment of photoelectrons, forming electron-electron repulsion between nanographene layers. In fact, when the electron repulsion between graphene layers reaches a certain high level, it will be sufficient to cause the separation of the graphene layers[31,32]. The adverse effects of defects in nanographene are eliminated by the saturation with excess electrons, and their electron-trapping ability contributes to the formation of the repulsive electric field. As a result, PESL exhibits an excellent ability to resist interference from humid atmospheric conditions.

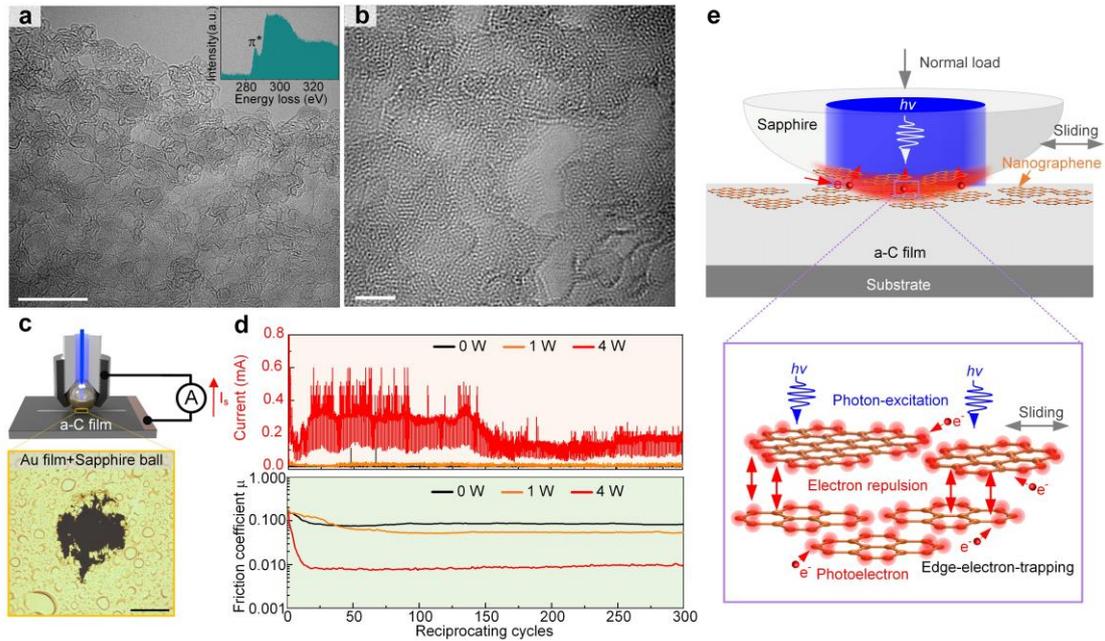

**Fig. 3 | Formations of nanographene and photoelectron. a** and **b,** TEM images of the transfer film on the sapphire ball. Parallel-stacked nanographene is observed. Inset is the EELS. **c,** Schematic illustration of electric current measurement during PESL test. An Au film was sputtered on the sapphire ball to make the circuit conductive. And after a pre-run-in 60 s, the Au film on the friction contact area was worn off to ensure light transmission of the ball. **d,** Current and friction curves of PESL test under different laser powers. PESL is accompanied by an order-of-magnitude increase in the number of electrons at the friction interface. **e**, Schematic illustration of PESL mechanism. PESL originates from the laser-irradiating induced formation of nanographene-layered interface and photoelectrons enriched at the interface, forming a repulsive electric field between the nanographene layers. Scale bars, 10 nm (**a**), 2 nm (**b**), 100 μm (**c**).

The above findings confirm the viability of our strategy for the in-situ fabricating and control of PESL. This strategy has also been successfully verified with a-C:H films. Compared with a-C films, a-C:H films typically exhibit ultra-low friction in inert environments due to the hydrogen passivation effect; however, they are more susceptible to atmospheric disturbances. Our experimental results show that a-C:H films also exhibited PESL characteristics in atmospheric environment along with excellent real-time controllability (Supplementary Fig. S8).

Moreover, we have experimented with changing the incident laser from violet light ($\lambda$=450 nm, $E_{450\ nm}$ =2.75 eV) to infrared light ($\lambda$=808 nm, $E_{808\ nm}$ =1.53 eV). Then, the photon energy decreases, but the photon penetration is enhanced, and the capacity to generate photoelectrons at the interface is reduced, resulting in a slight increase in the friction coefficient ($u$~0.010) (Supplementary Fig. S9). This observation further underscores the key role of photoelectrons in achieving PESL.

In conclusion, we discovered the phenomenon of PESL and demonstrated the feasibility of in-situ fabrication and control of PESL, which exhibits high robustness and rapid response in atmospheric environment. This remarkable performance is attributed to the laser-irradiating induced formation of nanographene-layered interface and enrichment of photoelectrons at the interface, forming considerable electron repulsion between the nanographene layers. The incorporation of electron repulsion eliminates the impediment caused by structural defects on the interlayer shearing of nanographene. Our strategy of in-situ fabricating superlubric surfaces, enhanced by electron repulsion, opens a new way for achieving superlubricity, and offers novel insights for smart friction, mechanical motion control and light manipulation.

(2012).

34. Cheng, C. et al. Friction-induced rapid restructuring of graphene cap layer at sliding surfaces: short run-in period, *Carbon* **130**, 215-221 (2018).

**Methods**

**Sample preparations**

The amorphous carbon (a-C) films were fabricated by using an electron cyclotron resonance (ECR) plasma sputtering system. The detailed description of the sputtering system was reported in our previous works[33, 34]. The a-C films were deposited on silicon substrates (p-type <100>, resistivity: <0.0015 Ω·cm) and 300 nm $SiO_2$/Si substrates with ion irradiation in divergent electron cyclotron resonance (DECR) plasma sputtering type. The background pressure of vacuum chamber was $8×10^{-5}$ Pa and the argon working pressure was $1×10^{-1}$ Pa. During the film deposition, ion irradiation was realized with a substrate bias voltage of -10 V. The film deposition rate was 3 nm/min. The thicknesses of a-C film/Si substrate and a-C film/300 nm $SiO_2$/Si substrate were 500 nm and 100 nm, respectively. Hydrogenated amorphous carbon (a-C:H) films with 500 nm in thickness were fabricated on Si substrates (Lirenaibang Coating Technology, Dongguan, China.)

**Frictional tests**

Friction tests of amorphous carbon films sliding against sapphire balls with in-situ laser irradiation were performed using a customized ball-on-plate tribometer, as shown in Supplementary Fig. S1. A laser was emitted from the laser source, passing through a zoom lens, a hollow weight, and a hollow ball holder, before penetrating a transparent sapphire ball to achieve in-situ irradiation of the frictional contact interfaces. The zoom lens was mounted on a four-axis lens mount and connected to the laser source via a lens bracket, which allows for laser focusing and spot size adjustment. The sapphire ball was fixed into the ball holder, which comprises a stainless-steel sleeve and a polyformaldehyde base. The polyformaldehyde base could insulate the heat conduction from the laser-induced temperature rise to the cantilever. The hollow weight was used to apply the normal load. The friction force was detected with strain gauges that were attached to the cantilever. The test accuracy for the friction force was 0.01 N. The normal load was applied from 3 N to 13 N. The sliding velocity was 20 mm/s, and the stroke length was 20 mm. The reciprocating cycle was defined as the complete movement of the sample slide from one end of its stroke to the other and back to the starting position. The tests were conducted in a clean room with a temperature of 24 °C and a relative humidity of 50-70%. Each test was repeated more than five times.

Supplementary Fig. S2 illustrates some basic characteristics of the friction tests with in-situ laser irradiation. Before test, the actual laser power was calibrated using an optical power meter. The laser spot was detected with a laser spot measuring instrument. As shown in Supplementary Fig. S2a, the diameter of laser spot was about 1~2 mm, which increases with the increase of laser power. An infrared thermal imager was employed to measure the temperature within the friction test region. From Supplementary Fig. S2b, it can be observed that the maximum temperature was located near the ball, reaching a maximum of 69.7 °C after 1000 reciprocating cycles of friction under the 4 W superlubricity condition. While the cantilever, which measures the friction force, remained at room temperature. This indicated that the laser-induced heating did not conduct to the cantilever beam, nor did it interfere with the friction force measurement, as verified in Supplementary Fig. S2c.

**Electric current measurement during PESL test**

First, an Au film was sputtered on the sapphire ball using a sputter coater (Cressington sputter coater 108 auto) to make the circuit conductive. And after a pre-run-in 60 s, the Au film on the friction contact area was worn off to ensure light transmission of the ball. Currents was measured with a low noise current preamplifier (SR570, Stanford Research System).

**Characterization techniques**

The transfer films on sapphire balls and wear tracks were imaged with an optical microscope (Nikon Eclipse LV150 N). The cross-sectional profiles of wear tracks were measured with a profilometer (Bruker, Dektak-XT). The surface morphology of a-C film and wear track were measured by atomic force microscopy (AFM, Bruker, Innova) in a tapping mode. Silicon tip with curvature radius of 4 nm was used, and the images were obtained with a scan size of 5 μm×5 μm and a scan frequency of 0.8 Hz.

The nanostructures of the transfer films and wear tracks were analyzed with a Raman spectroscope (Horiba, HR-Resolution; wavelength of 532 nm) and a Cs-corrected transmission electron microscope (TEM, Thermo Fisher Scientific, Titan3 Cubed Themis G2 300). The TEM was operated at 80 kV to avoid possible damages or recrystallization caused by electron irradiation.

**TEM sample preparation**

The plan view TEM sample of the transfer film was prepared by a custom-made micron-precision positioning table equipped with a microscope and a microprobe station (with a tungsten microprobe of 1 mm tip diameter). The process of the TEM samples fabrication was presented in Supplementary Fig. S3. The target regions were first confirmed under the microscope, then the microprobe was conducted to approach the target regions and scratch the surface with a gentle force to make the micro scrapes of the transfer film adhere to the probe. After that, a droplet of acetone was dropped on a cooper grid, then the probe adhered with transfer film was conducted to touch the grid and slightly scratch to make scrapes adhere to the grid to finish sample preparation.

The plan view TEM sample of the wear track was prepared with a wetting transfer method, as shown in Supplementary Fig. S4. To ensure the sample thin enough, a 100 nm-thick a-C film was deposited on a 300 nm $SiO_2$/Si substrate. The film was firstly subjected to normal friction sliding without laser for 20 minutes to further reduce the thickness. Subsequently, a 450 nm laser was turned on to induce a stable superlubric state with the a-C film. After maintaining the superlubric state for 10 minutes, the friction test was terminated. Under the assistance of an optical microscope, a 2 mm×2 mm sample containing the wear track was cut out from the film using a diamond pencil.

Then the sample was coated with a thin polymethyl methacrylate (PMMA) layer on the spin coater with a rotation rate of 2,000 round per minute and then heated for 3 min with 120 °C. After which, the sample was immersed in 10% hydrofluoric acid (HF) solute for 3~4 h to dissolve the $SiO_2$ layer and completely separated from the substrate. Afterward, the sample was transferred to a cooper grid in pure water and cleaned with acetone drops for 5 min to dissolve the PMMA layer completely to obtain the pure film sample.

**Acknowledgements**

The authors would like to acknowledge the National Natural Science Foundation of China (Nos. 52175179), Natural Science Foundation of Guangdong Province (No. 2022A1515010555), and Shenzhen Fundamental Research Program (No. JCYJ20210324100406018). The Electron Microscopy Center (EMC) of Shenzhen University is kindly acknowledged for the technical supports in TEM and FIB.

**Author contributions**

C.C. and D.D. designed and supervised the research. Z.Z., F.L., and H.W. performed the experiments and analyzed the data. P.X. assisted with all experiments. C.C. and D.D. co-wrote the paper. All authors discussed the results and commented on the manuscript.

**Competing interests**

The authors declare no competing interests.